\def\preprint{1}			

\ifdefined\preprint
  \documentclass[preprint,review,12pt]{elsarticle}
\fi
\ifdefined\wordcount
  \documentclass[final,3p,times,twocolumn]{elsarticle}
\fi
\ifdefined\final
  \documentclass[final,3p,times,twocolumn]{elsarticle}
\fi

\usepackage{graphicx,stfloats}
\usepackage{color}

\usepackage{amsmath}
\usepackage{tabularx}
\ifdefined\preprint
\usepackage[letterpaper,margin=1.in]{geometry}
\fi
 \usepackage{float}

\biboptions{sort&compress}

\journal{Combustion and Flame}

\begin{document}

\begin{frontmatter}

\title{Ignition criteria and the effect of boundary layers on wedge-stabilized oblique detonation waves}

\author[fir]{Christian L. Bachman\corref{cor1}}
\ead{christian.bachman@nrl.navy.mil}

\author[fir]{Gabriel B. Goodwin}

\address[fir]{Naval Research Laboratory, Washington, DC 20375, United States}
\cortext[cor1]{Corresponding author:}

\begin{abstract}

Simulations of a supersonic, premixed, reacting flow over a wedge were performed to investigate the effect of a boundary layer on the wedge surface on ignition and stability of oblique detonation waves (ODWs). Two computational domains were used: one containing a wedge of a single angle with a straight after-body, and the other containing a double-angle wedge geometry. Both domains were channels with a supersonic inflow of stoichiometric hydrogen-air and a nonreflecting outflow, and the wedge was modeled using an immersed boundary method. The compressible reactive Navier-Stokes equations were solved using a high-order numerical algorithm on an adapting grid. Inviscid and viscous wedge surfaces were modeled using slip and no-slip adiabatic boundary conditions, respectively. Inviscid wedge surface cases are presented for a range of inflow conditions and compared to previous work outlining several different ODW structures. An ignition criterion is established as an accurate method of predicting the formation of an ODW for a given inflow temperature, Mach number, wedge angle, and length of the inviscid wedge surface. A viscous wedge surface is then considered for a Mach 5 inflow at temperatures of 600 K, 700 K, and 800 K. The 600 K flow ignites in the boundary layer, but does not detonate, while the 700 K and 800 K flows ignite and form ODWs. It was determined that ODW formation depends on the degree of augmentation of the leading oblique shock wave by the burning boundary layer and that ODW formation is therefore predictable based on the ignition criterion. The 700 K flow produces a unique oscillatory mode of a receding ODW followed by a redetonation event occurring near the leading edge. A mechanism for this cycle, which repeats indefinitely, is proposed.

\end{abstract}


\end{frontmatter}

\ifdefined \wordcount
\clearpage
\fi

\section{Introduction}
\label{Introduction}

	A detonation is a pressure-driven combustion wave which travels at supersonic speeds in a premixed, chemically reactive background gas \cite{glassman2014combustion}. Detonations are a promising combustion mechanism for supersonic and hypersonic airbreathing propulsion because of the thermodynamic work achieved by the pressure gain and the reduced size of detonation engines. For these reasons many decades of airbreathing propulsion has focused heavily on pulsed detonation engines (PDEs) \cite{kailasanath2003recent}, rotating detonation engines (RDEs) \cite{rankin2017chemiluminescence}, and oblique detonation wave engines (ODWEs) \cite{cambier1990numerical}. ODWEs most resemble the typical scramjet engine configuration in that the isolator and combustor are often shortened to a single turning angle after fuel injection, on which an oblique detonation wave (ODW) is anchored. 

	Analyses have shown that the ODWE has comparable performance to the scramjet engine, with the main benefits of reduced engine size and increased thrust produced through pressure-gain combustion \cite{ashford1996oblique}. This has prompted analytical \cite{pratt1991morphology}, experimental \cite{morris1996plif, morris1998shock}, and computational \cite{glenn1988numerical, li1994detonation, silva2000stabilization} work to understand the initiation and stabilization of ODWs on anchoring bodies. Previous studies have looked at blunt bodies and wedges, both as projectiles through reactive mixtures \cite{lefebvre1995numerical, maeda2013initiation, li1995dynamics}, and as stationary perturbations within premixed hypersonic flows \cite{kamel1996plif, morris1998shock, walter2006numerical}. In an ODWE the wedge is preferable due to the lesser total pressure loss through an oblique detonation than through the partially normal detonation wave produced by blunt bodies.

	A majority of the work pertaining to wedge-stabilized ODWs is computational, due to experimental difficulty in obtaining the high flow enthalpy, reactivity, and Mach numbers required to ignite a detonation and stabilize an ODW. Many of these computational studies ignore the effects of boundary layers on the wedge surface with an inviscid, slip wall assumption. This is motivated by an early study by Li et al. \cite{li1993effects} that reported minor differences in the ODW structure when boundary layers are present. A more recent study by Fang et al. \cite{fang2019effects} considered a viscous wedge surface for two different Mach numbers and found that the boundary layer has a non-negligible effect on the ODW structure in the low Mach number case.

	This paper adds to the growing body of work on high-fidelity simulations of wedge-stabilized ODWs in hypersonic hydrogen-air mixtures. We further consider the effect of boundary layers on the ODW structure and the importance of modeling a viscous wedge surface interaction. A particular transient case is presented in which an oscillatory, quasi-stable ODW results from the presence of a boundary layer and its interaction with the ODW.

\section{Background}
\label{Background}

	An analysis of ODW stability was performed by Pratt et al. \cite{pratt1991morphology} using oblique shock polars with chemical heat release. A nondimensional heat release defined as $\tilde{Q} = Q/C_p T$, where $Q$ is the heat release, $T$ is the inflow static temperature, and $C_p$ is the constant pressure heat capacity, leads to a Chapman-Jouguet (CJ) ODW with CJ wave angle, $\beta_{CJ}$,  corresponding to a CJ Mach number, $M_{CJ}$, and CJ deflection angle, $\theta_{CJ}$. These quantities are defined as

\begin{equation}
	M_{CJ}^2 = [1 + \tilde{Q}(\gamma + 1)] + \sqrt{[1 + \tilde{Q}(\gamma + 1)]^2 - 1},
\end{equation}

\begin{equation}
	\beta_{CJ} = sin^{-1}(M_{CJ}/M),
\end{equation}

\noindent and

\begin{equation}
	\theta_{CJ} = \beta_{CJ} - tan^{-1}\left( \frac{1+\gamma M_{CJ}^2}{(\gamma+1)M_{CJ}^2\sqrt{(M/M_{CJ})^2-1}} \right),
\end{equation}

\noindent where $\gamma$ is the ratio of specific heats and $M$ is the inflow Mach number. The quantity $M_{CJ}$ is simply the Mach number of a normal CJ detonation wave, and $\beta_{CJ}$ is defined such that the component of the inflow Mach number that is normal to a CJ-ODW is $M_{CJ}$ and the resulting flow deflection angle is $\theta_{CJ}$.

\begin{figure}[h]
		\centering
		\includegraphics[width=88mm]{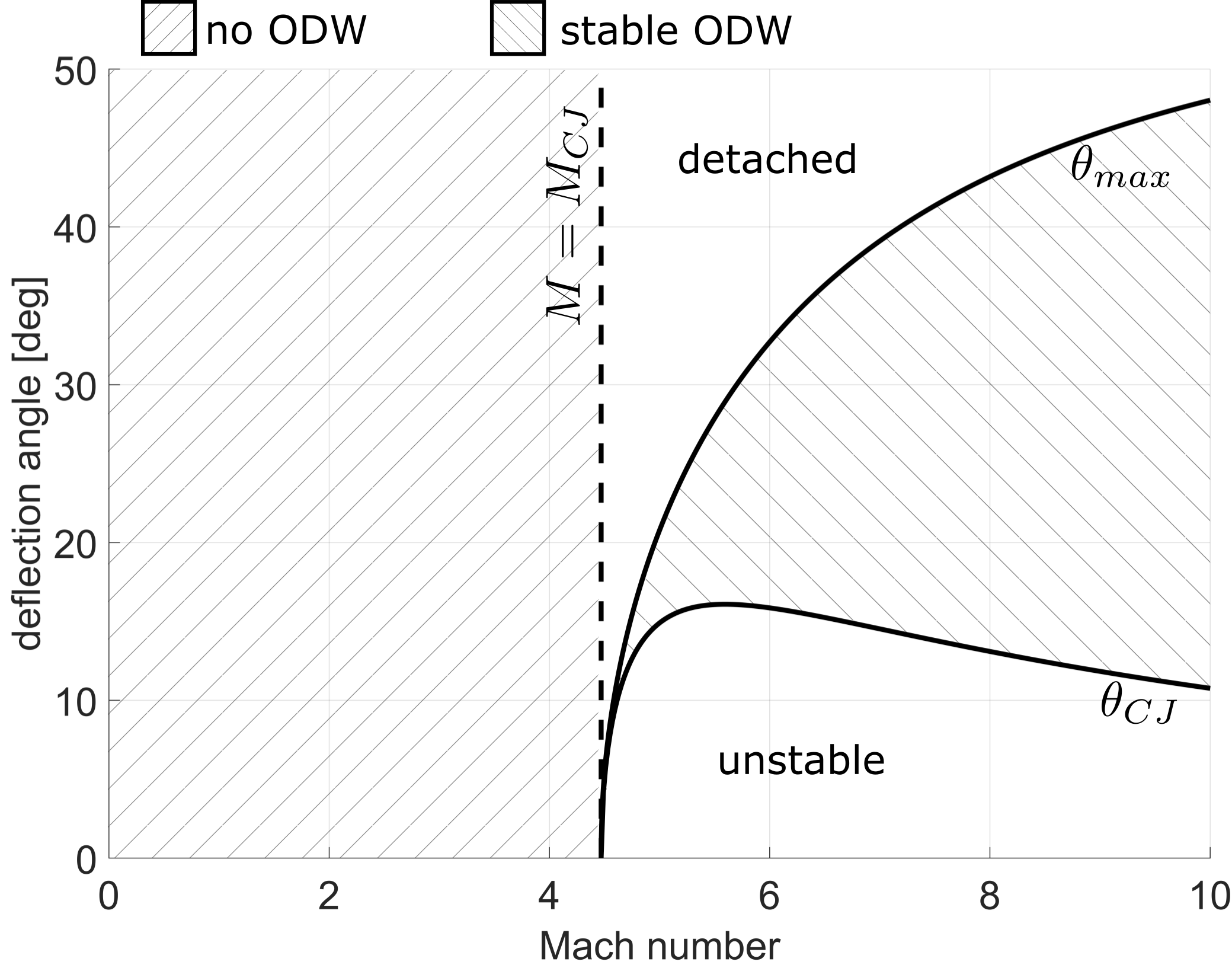}
		\caption{ODW stability limits for stoichiometric hydrogen-air at 500 K.}
		\label{fig:inv_stab}
	\end{figure}

	The reactive shock polar analysis also leads to a deflection angle at which the ODW becomes detached from the leading edge, analogous to adiabatic oblique shock wave (OSW) detachment. This maximum deflection angle, $\theta_{max}$, is less than that for an OSW and can be plotted with $\theta_{CJ}$ against inflow Mach number, as shown in Fig.~\ref{fig:inv_stab} for a 500 K inflow. In order for a flow to support an ODW, $M$ must be greater than $M_{CJ}$. Furthermore, stability is predicted for deflection angles ranging from $\theta_{CJ}$ to $\theta_{max}$, with the resulting ODWs ranging from CJ-ODWs to overdriven ODWs, respectively. This analysis enables the prediction of stability limits for a given mixture of gas at a given inflow static temperature and pressure, provided a value for $\tilde{Q}$ can be determined by evaluation of some chemical kinetic model. 

\begin{figure}[h]
		\centering
		\includegraphics[width=88mm]{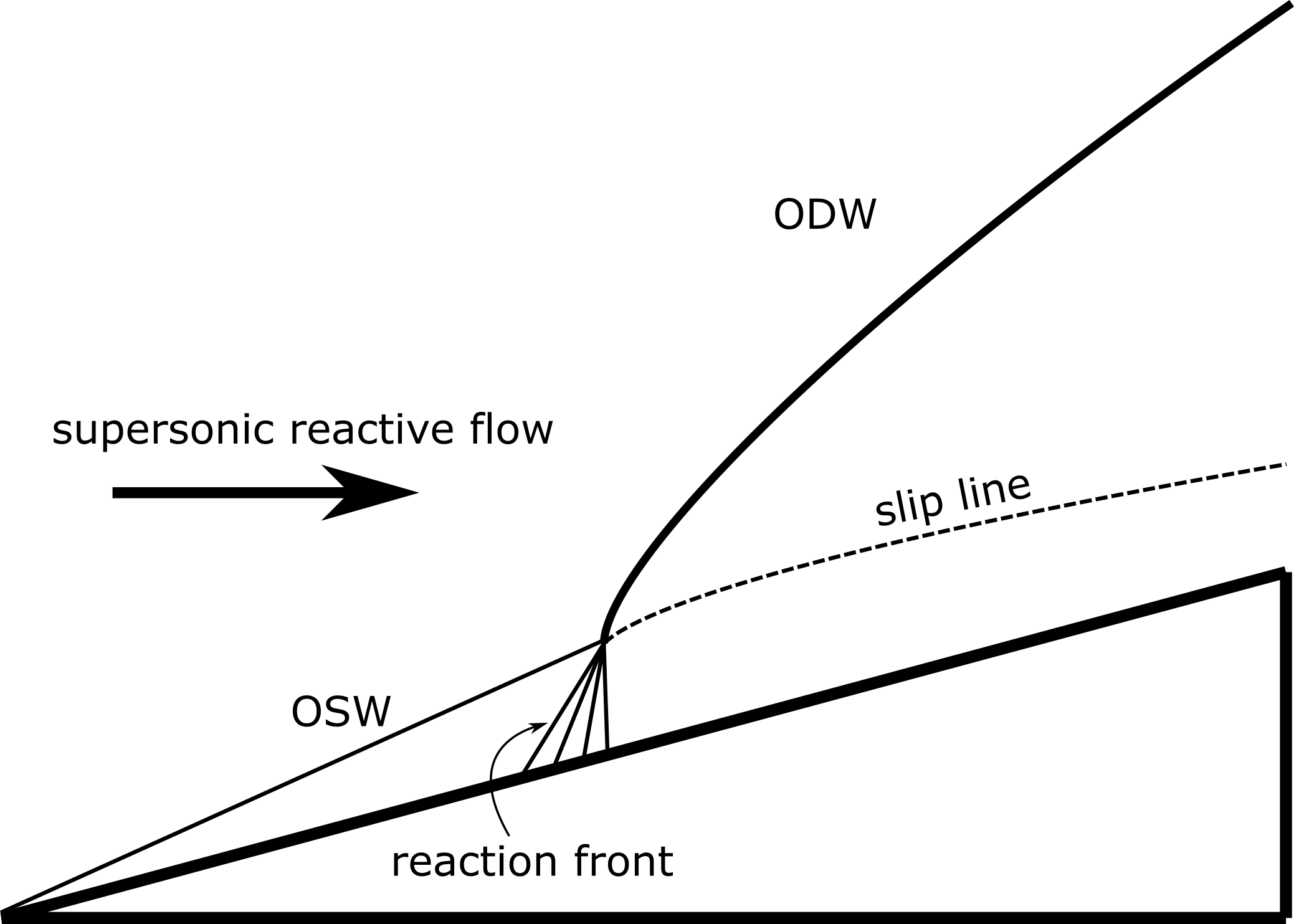}
		\caption{Flowfield structure of an ODW on an inviscid wedge surface.}
		\label{fig:inv_struct}
	\end{figure}

	Some of the first computational studies by Glenn and Pratt \cite{glenn1988numerical} and Li et al. \cite{li1994detonation} began to validate the stability of the shaded region between the two limits. The latter study was the first to establish the typical structure of a wedge-stabilized ODW. This structure is shown in Fig.~\ref{fig:inv_struct} and is characterized by an OSW, behind which is an induction region that terminates some distance downstream in a reaction front. The intersection of the OSW with the reaction front steepens the OSW into an overdriven ODW, the angle of which relaxes in the farfield to $\beta_{CJ}$. The streamlines passing through the induction region and ODW are separated by a slip line through the burned gas. The length of the induction region corresponds to the ignition delay time of the mixture behind the OSW.

	Follow-on work by Li et al. \cite{li1993effects} considered a viscous wedge surface ODW, that is a no-slip condition on the wedge surface, to determine the effect of boundary layers on the ODW structure. The study reported that the overall structure remained the same, apart from a thin ignited boundary layer resulting in a slightly steeper OSW and smaller induction region than in the inviscid wedge surface case with a slip wall condition along the wedge. Notable experimental work by Morris et al. \cite{morris1996plif,morris1998shock} used an expansion tube to produce ODWs on a wedge in different dilutions of hydrogen-oxygen mixtures, reporting transient wedge surface pressure data and instantaneous planar laser-induced fluorescence (PLIF) and schlieren images for each run. These studies reported smaller induction lengths than predicted by analytical models, but dismissed the presence of a boundary layer as a potential reason for this discrepancy. 

	Subsequent numerical work has used these findings to justify the use of an inviscid wedge surface when studying ODW stability. Figueira da Silva and Deshaies \cite{silva2000stabilization} and Walter and Figuiera da Silva \cite{walter2006numerical} confirmed that the induction region length corresponds to ignition delay time and found that the transition from OSW to ODW can have a smooth or abrupt structure. They pointed out that induction lengths in experiments are shorter than computed and suggested that it may be due to the inviscid wedge surface in the computations. Choi et al. \cite{choi2009unstable} simulated the experimental conditions of Morris et al. and found that the particular cases with a wedge angle exceeding the detachment angle are transient and oscillatory. Teng et al. \cite{teng2014numerical} also found a transient ODW structure for a low Mach number inflow, and noted different induction zone structures depending on Mach number, labeling them $\lambda$, X and Y shaped, where the $\lambda$ structure resembles the abrupt OSW-to-ODW transition. Liu et al. \cite{liu2015analytical} computed another possible transient state where the ODW structure propagates upstream and stabilizes as a prompt ODW anchored to the leading edge, which may occur for relatively low Mach numbers and large wedge angles. The smooth and abrupt transitions from OSW to ODW were then shown by Liu et al. \cite{liu2016structure} and Teng et al. \cite{teng2017initiation} to occur for high and low inflow Mach numbers, respectively. 

	The existing literature does much to illuminate the possible structures and stability of ODWs, demonstrating a high sensitivity to Mach number and wedge angle. The potential importance of a boundary layer in ODW formation and stability, however, has not received as much attention. A recent study by Fang et al. \cite{fang2019effects} considered ODWs on a viscous wedge surface in Mach 10 and 7 flows and found that the smooth OSW-to-ODW transitions at higher Mach numbers are negligibly affected by the boundary layer, but that lower Mach number cases with abrupt transitions are greatly affected. A transient state driven by the boundary layer was found.

	In light of the quantity of numerical work using inviscid wedge surfaces, and the scarcity of viscous wedge surface studies that take advantage of modern computing power, the present paper first considers ODWs on inviscid wedge surfaces in comparison to the cited works as validation of the numerical model and further confirmation of the nature of ODW stability, then includes a viscous boundary condition on the wedge surface to look at the effects of boundary layers across a range of inflow conditions and wedge angles. 

\section{Numerical Model}
\label{Numerical Model}

The simulations solve the Navier-Stokes equations and ideal gas equation of state for a compressible, unsteady, chemically reacting gas,  

\begin{equation} \label{eqn1}
\frac{\partial \rho}{\partial t} + \nabla \cdot(\rho \mathbf{U}) = 0
\end{equation}
\begin{equation}
\frac{\partial (\rho\mathbf{U})}{\partial t} + \nabla \cdot(\rho\mathbf{UU}) + \nabla P = \nabla\cdot\mathbf{\hat{\tau}} 
\end{equation}
\begin{equation}
\frac{\partial E}{\partial t} + \nabla\cdot((E + P)\mathbf{U}) = \nabla\cdot(\mathbf{U}\cdot\hat{\mathbf{\tau}}) + \nabla\cdot(K\nabla T) - \rho q\dot{w} 
\end{equation}
\begin{equation}
\frac{\partial (\rho Y)}{\partial t} + \nabla\cdot(\rho Y\mathbf{U}) + \nabla\cdot(\rho D\nabla Y) - \rho\dot{w} = 0
\end{equation}
\begin{equation}
P = \frac{\rho RT}{M}
\end{equation}

\noindent where $\rho$ is the mass density, \textbf{U} is the fluid velocity, $E$ is the energy density, $P$ is the static pressure, $T$ is the static temperature, $Y$ is the mass fraction of reactant, $\dot{w}$ is the rate of reaction, $q$ is the chemical heat release, $K$ is the thermal conduction coefficient, $D$ is the mass diffusion coefficient, $R$ is the universal gas constant, and $M$ is the molecular weight. The viscous stress tensor is given by

\begin{equation}
\mathbf{\hat{\tau}} = \rho\nu((\nabla\mathbf{U}) + (\nabla\mathbf{U})^T-\frac{2}{3}(\nabla\cdot\mathbf{U})\mathbf{I}),
\end{equation}

\noindent where \textbf{I} is the unit tensor, $\nu$ is kinematic viscosity, and the superscript $T$ indicates matrix transposition. Conversion of reactant to product is modeled by Arrhenius kinetics,

\begin{equation}
dY/dt \equiv \dot{w} = -A\rho Y\exp(-E_\mathrm{a}/RT)
\end{equation}

\noindent where $A$ is the pre-exponential factor and $E_\mathrm{a}$ is the activation energy. Viscosity, mass diffusion, and thermal conduction depend on local temperature and density,

\begin{equation} \label{eqn2}
\nu = \nu_0\frac{T^n}{\rho},   D = D_0\frac{T^n}{\rho},   \frac{K}{\rho C_p} = \kappa_0\frac{T^n}{\rho},
\end{equation}

\noindent where $\nu_0$, $D_0$, and $\kappa_0$ are calibrated constants, and $n$ = 0.7 enforces the temperature dependence of the system. 

	Equations~(\ref{eqn1})--(\ref{eqn2}) are solved using a fifth-order accurate Godunov scheme with third-order Runge-Kutta time integration \cite{houim2011low} and locally adapting grid refinement \cite{zhang2016boxlib}. Wedge surfaces are modeled using the immersed boundary method [IBM] \cite{chaudhuri2011use}. Reaction and transport parameters throughout the governing equations are calibrated to quantitatively reproduce ignition, flame, and detonation properties, calculated using detailed chemistry, for a stoichiometric hydrogen-air mixture. The parameters are calibrated using a genetic algorithm optimization procedure \cite{kaplan2019chemical}. Table~\ref{table:params} shows the values of the calibrated input parameters and the target output parameters. This reaction model quantitatively reproduces flame acceleration \cite{gamezoCNF2008, ogawaJLPP2013, GoodwinCNF2016, kesslerCNF2010}, onset of turbulence \cite{houim2016role, poludnenko2015pulsating}, and various mechanisms of deflagration-to-detonation transition (DDT) observed in experiments \cite{xiaoflame, xiao2015formation, OranCNFrev2007}, and has been used to study supersonic boundary layer combustion \cite{goodwin2018premixed}. 

\begin{table}[h]
	\caption{Calibrated model parameters and reproduced combustion wave properties for stoichiometric hydrogen and air.}
	\centering
	\ifdefined\preprint
		\footnotesize
	\fi
	\ifdefined\wordcount
		\small
	\fi
		\begin{tabularx}{140mm}{p{1.6cm}p{4.77cm}p{7.63cm}}
		\hline
		Input\\
		\hline	 	

		$\gamma$ & 1.19 & Adiabatic Index \\
		$M$ & 24.5 g/mol & Molecular Weight \\
		$A$ & 4.27 $\times$ $10^{13}$ cm$^3$/(g s) & Pre-Exponential Factor \\
		$E_\mathrm{a}$ & 1.58 $\times$ $10^{5}$ J/mol & Activation Energy \\
		$q$ & 4.38 $\times$ $10^{6}$ J/kg & Chemical Energy Release \\
		$\nu_\mathrm{o}$ & 1.7 $\times$ $10^{-6}$ g/s-cm-K$^{0.7}$ & Viscosity \\
		$\kappa_\mathrm{o}$, $D_\mathrm{o}$ & 2.9 $\times$ 10$^{-6}$ g/s-cm-K$^{0.7}$ & Transport Constants \\

		\hline
		Output\\
		\hline

		$S_\mathrm{L}$ & 227.2 cm/s & Laminar Flame Speed \\
		$T_\mathrm{b}$ & 2379 K &  Adiabatic Flame Temperature \\
		$T_\mathrm{CV}$ & 2748 K &  Isochoric Flame Temperature \\
		$x_\mathrm{l}$ & 3.31 $\times$ $10^{-2}$ cm & Laminar Flame Thickness \\
		$D_\mathrm{CJ}$ & 1.97 $\times$ $10^5$ cm/s & CJ Detonation Velocity \\
		$x_\mathrm{d}$ & 2.45 $\times$ $10^{-2}$ cm & Half-Reaction Thickness \\

		\hline 		
	\end{tabularx}
	\label{table:params}
\end{table}

	The calculations described below used computational grids with coarsest size $dx_{max}$ = 1.4 mm and minimum size $dx_{min}$ = 22 $\mu$m. Dynamic grid refinement is performed in regions of high gradients in density, pressure, and reactant mass fraction, and across boundary layers when present. The minimum cell size corresponds to 11.2 cells across the half-reaction thickness and 15.1 cells across the laminar flame thickness in Table~\ref{table:params}. A resolution study was performed to test the sensitivity of the system to minimum grid size. This investigation is discussed in detail in Sec.~\ref{resolution}.

\begin{figure}
	\centering
	\includegraphics[width=88mm]{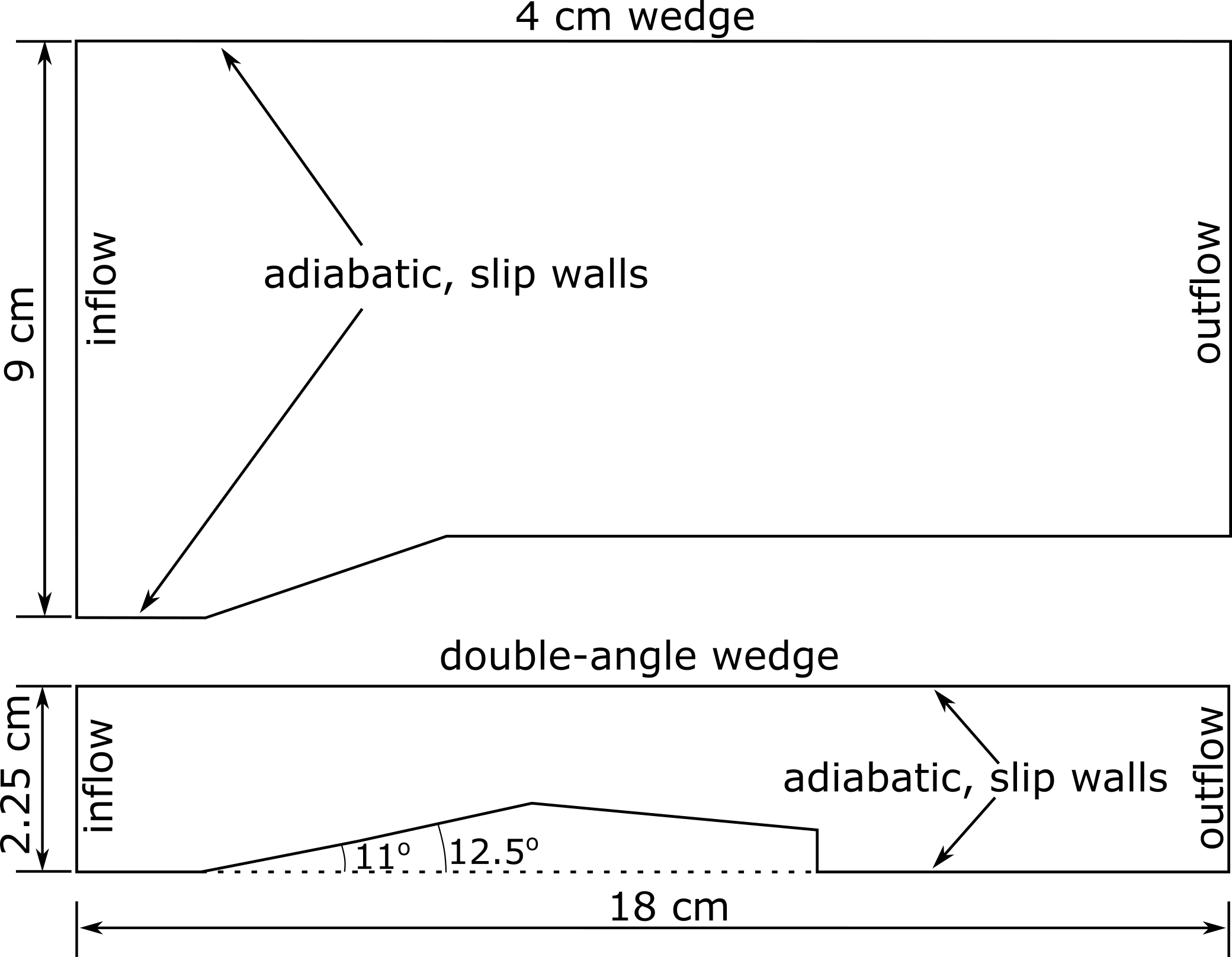}
	\caption{Schematics of computational domains 1 (top) and 2 (bottom).}
	\label{fig:domain}
\end{figure}

	Figure~\ref{fig:domain} shows the two computational domains that were used in this work. Both are rectangular domains of length 18 cm with a supersonic inflow of stoichiometric hydrogen-air on the left boundary, and a non-reflective outflow on the right. The upper and lower boundaries are adiabatic, slip walls. The wedges are placed in both domains 2 cm from the left boundary, and the IBM allows a slip or a no-slip condition to be imposed on the wedge surface, enabling inviscid or viscous wedge surface boundary conditions, respectively. Domain 1 has a height of 9 cm, and uses a wedge that is 4 cm in length. Various wedge deflection angles were considered in this domain. Domain 2 has a height of 2.25 cm and a double-angle wedge that is initially 11$^o$ and steepens 2 cm downstream to 12.5$^o$. Domain 2 models the upper half of the test section in the Mach 5 hypersonic combustor facility in the Center for Advanced Turbomachinery and Energy Research (CATER) at University of Central Florida (UCF), where the double-angle wedge was initially proposed for experimental testing of ODW stability in hypersonic hydrogen-air mixtures \cite{sosa2020shock}. The inflow static temperature and Mach number are varied in both domains throughout the study, and the inflow static pressure is kept at 1 atm for all cases.

\section{Results}

\subsection{Inviscid wedge surface}
\label{inviscid}

	A slip wedge surface is considered first to confirm the ability of the model to produce the flow structures and stability regimes that have been established in previous works. Inflow temperatures of 500 K and 800 K are considered for domains 1 and 2, respectively. Calculations performed in domain 1 vary in inflow Mach number and wedge deflection angle. All of these cases lie within the stable ODW region, mapped in Fig.~\ref{fig:inv_stab}, for a 500 K inflow. 

\subsubsection{500 K Inflow}

\begin{figure}[h]
		\centering
		\includegraphics[width=88mm]{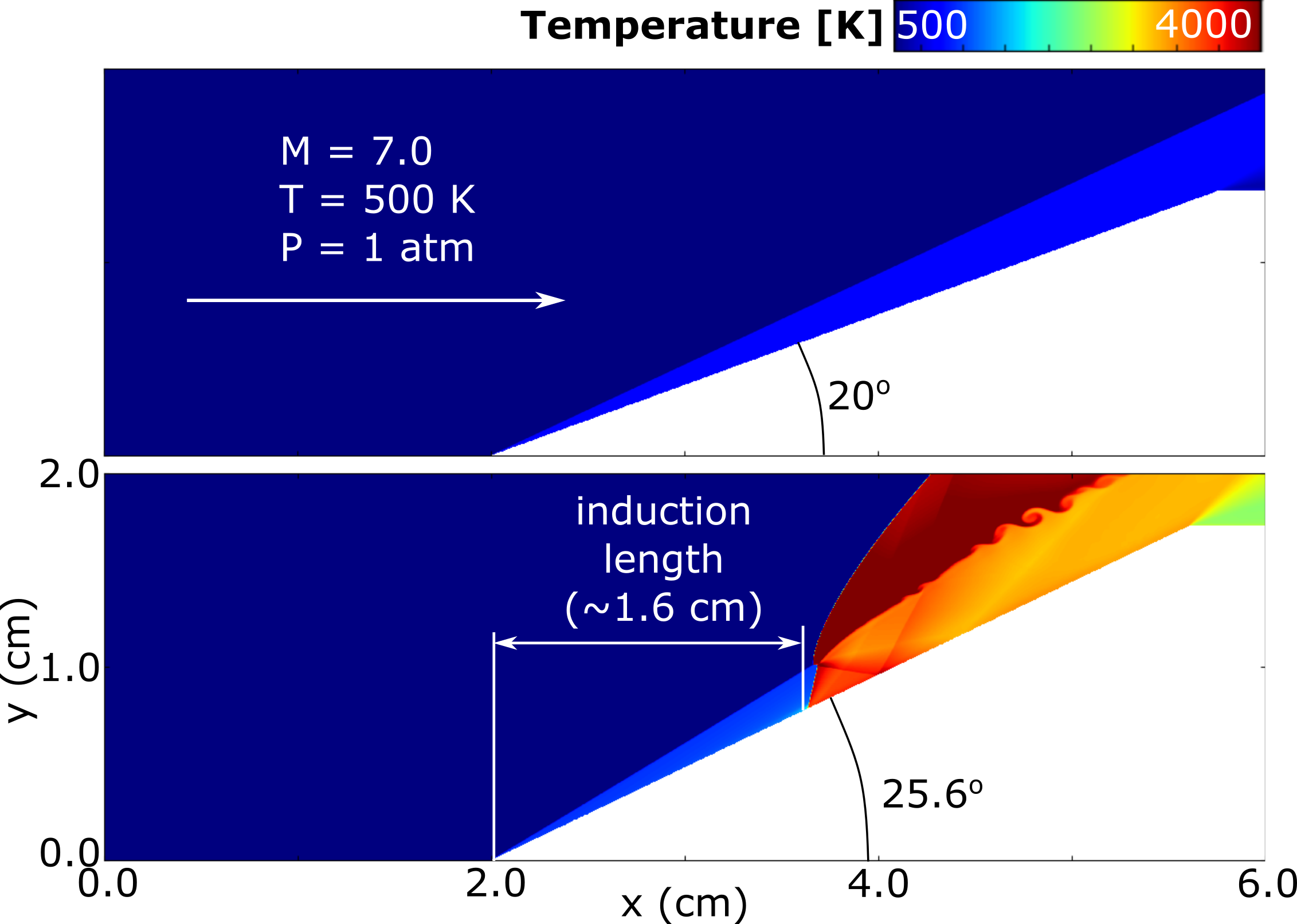}
		\caption{Mach 7 cases with 500 K, 1 atm inflow for 20$^o$ (top) and 25.6$^o$ (bottom) wedges of length 4 cm.}
		\label{fig:m7_odw}
	\end{figure}

	Figure~\ref{fig:m7_odw} shows the steady-state temperature for two Mach 7 cases with wedge angles of 20$^o$ and 25.6$^o$. Despite the flow conditions being within the stable ODW region shown in Fig.~\ref{fig:inv_stab}, the lesser angle does not provide enough shock compression to ignite the mixture, and the steady state is a non-reacting OSW. The flow does ignite on the 25.6$^o$ wedge surface, and the typical ODW structure is established.  

\begin{figure}[h]
		\centering
		\includegraphics[width=88mm]{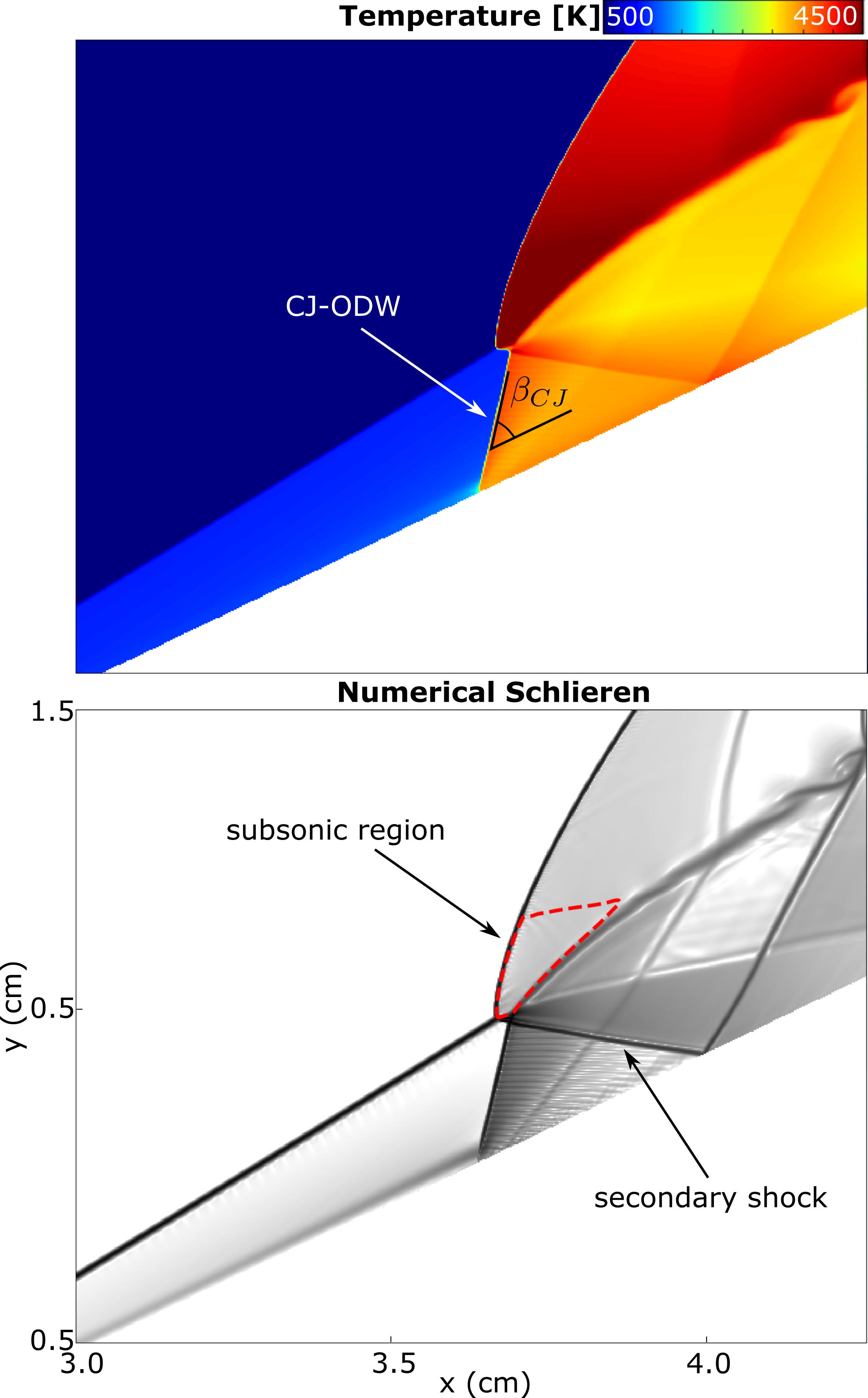}
		\caption{Stable ODW structure for Mach 7, 500 K inflow over a 25.6$^o$ wedge. The dashed line encloses the subsonic region.}
		\label{fig:m7_close}
	\end{figure}

	Figure~\ref{fig:m7_close} shows contours of temperature and numerical schlieren which provide a more detailed view of the ODW structure. The schlieren is overlaid by a red, dashed contour denoting the sonic line and encircling a subsonic region of the flowfield. The reaction front, which terminates the induction region, ignites along the wedge surface where the flow residence time of the shocked gas is greatest. As this combustion wave grows through the induction region, the reaction front becomes pressure driven and steepens into a detonation. The wave angle of this ODW terminating the induction region is equal to $\beta_{CJ}$ for the post-shock flow conditions. Thus, the reaction front is a CJ-ODW within the induction region, labeled as such in the figure. 

	This mechanism of wave strengthening has been noted before, and results in an abrupt OSW-to-ODW transition \cite{silva2000stabilization, teng2017initiation, fang2019effects}. The particular induction zone structure seen here matches the $\lambda$ shape noted by Teng et al. \cite{teng2014numerical} and is formed by the reflection of the terminating CJ-ODW from the OSW as a secondary shock, labeled in the schlieren of Fig.~\ref{fig:m7_close}. This leads to a train of shocks reflecting between the wedge surface and the slip line formed behind the ODW structure, seen in Figs.~\ref{fig:m7_odw} and \ref{fig:m7_close}. The intersection of the CJ-ODW with the initial OSW results in the formation of the overdriven ODW followed by the subsonic region. This ODW relaxes in the farfield to the wave angle $\beta_{CJ}$ corresponding to the freestream flow conditions.

\begin{figure*}
		\centering
		\includegraphics[width=144mm]{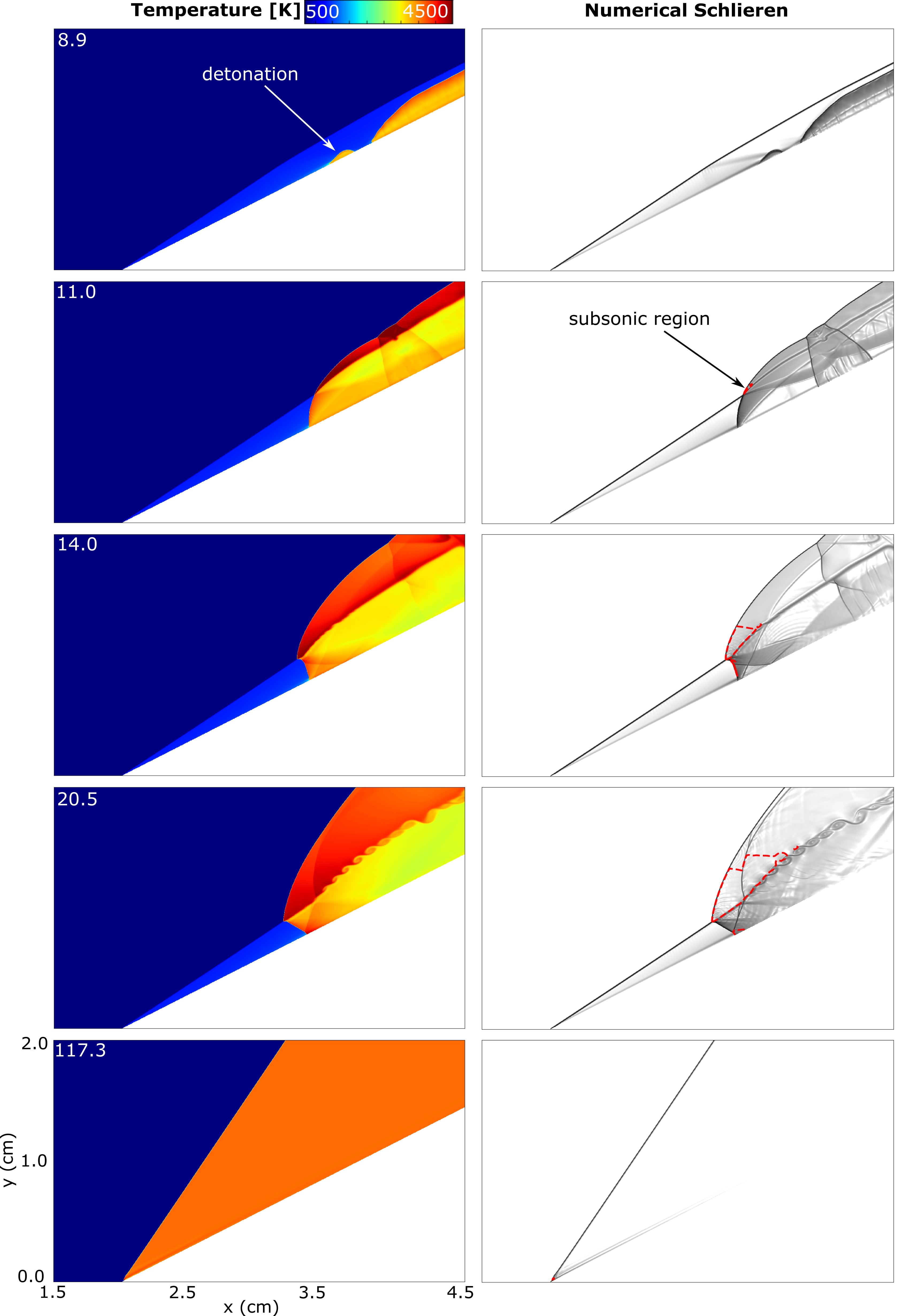}
		\caption{Time evolution of Mach 6, 500 K case with a 30$^o$ wedge of length 4 cm. The dashed line encloses the subsonic region, and timestamps are given in $\mu$s.}
		\label{fig:m6_t30_evol}
	\end{figure*}

	Figure~\ref{fig:m6_t30_evol} shows the time evolution of a transient structure with timestamps in $\mu$s in the top left corners of the temperature plots. This is a case with a 4 cm, 30$^o$ wedge and an inflow Mach number of 6, again at 500 K. The flow ignites along the wedge surface in the first frame at 8.9 $\mu$s. This flame front grows, strengthening into a detonation wave that intersects the OSW and leads to an ODW followed by a subsonic region. For this lower Mach number, however, the resulting $\lambda$ structure in the induction region is also partially overdriven, denoted by the sonic line at 14 $\mu$s. This leads to an unstable terminating wave that transitions to the Y shape structure seen at 20.5 $\mu$s, which has been noted to arise for low Mach numbers in previous work \cite{teng2014numerical}. The Mach stem forming the lower branch of the Y shape remains overdriven, and the entire structure propagates upstream before stabilizing as a prompt ODW anchored to the leading edge. This process is similar to those computed by Liu et al. \cite{liu2015analytical}. The wave angle of this prompt ODW is greater than the value of $\beta_{CJ}$ corresponding to the inflow conditions because the 30$^o$ deflection angle imposed by the wedge is much greater than $\theta_{CJ}$, as seen in Fig.~\ref{fig:inv_stab}. This results in an overdriven, but stable, ODW.

\subsubsection{800 K Inflow}

\begin{figure}
		\centering
		\includegraphics[width=88mm]{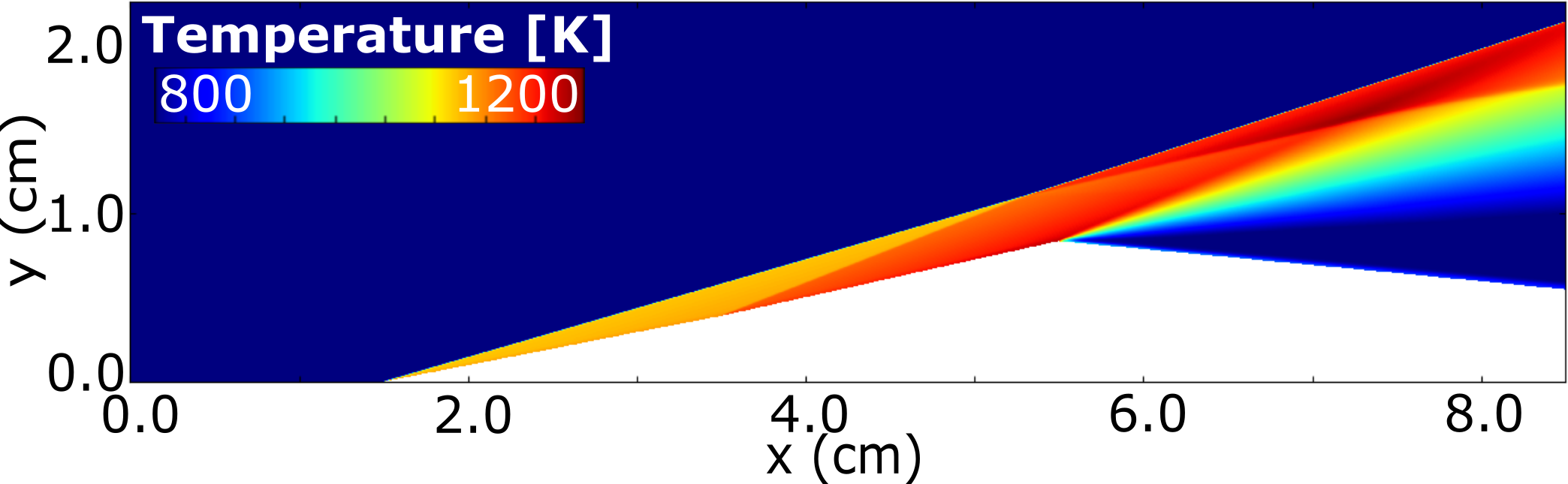}
		\caption{Steady-state temperature of the double-angle wedge in a Mach 7.3, 800 K flow. No ignition occurs.}
		\label{fig:ucf_m7}
	\end{figure}

	Two inviscid wedge surface cases were computed in domain 2 using the double-angle wedge. Mach numbers of 7.3 and 8.2 were used, each with an inflow temperature of 800 K. For an inflow Mach number of 7.3, the flow does not ignite and a distinct shock anchors to each of the two wedge angles. The steady state is shown in Fig.~\ref{fig:ucf_m7}. The two OSWs intersect and refract, continuing downstream and reflecting off the upper wall. Even with the post-shock temperature exceeding 1100 K the flow residence time is insufficient for shock-induced combustion to occur along the wedge surface before the wide turn angle downstream. This behavior is invariably linked to the ignition delay time of the post-shock conditions, discussed further in Sec.~\ref{ign crit}.

\begin{figure}[h]
		\centering
		\includegraphics[width=88mm]{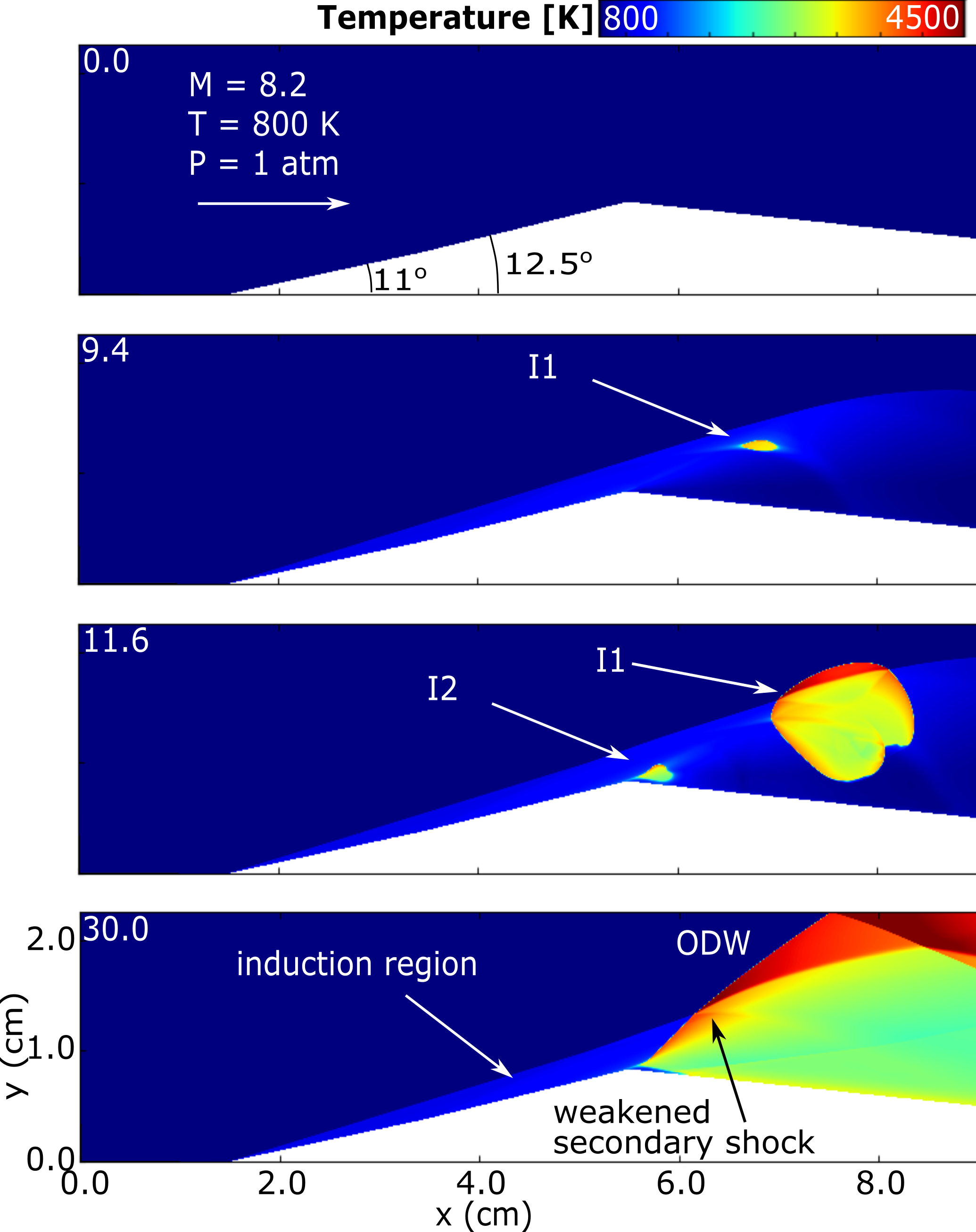}
		\caption{Double-angle wedge in a Mach 8.2, 800 K flow. Timestamps are given in $\mu$s.}
		\label{fig:double-angle_inviscid}
	\end{figure}

	With the higher Mach number of 8.2, two sources of ignition occur and lead to a stable ODW. The evolution of this case is shown in Fig.~\ref{fig:double-angle_inviscid}. The first ignition at 9.4 $\mu$s occurs at the intersection of the developing OSWs. The temperature and pressure produced is sufficient to autoignite the mixture, which causes a detonation wave to grow upward and downstream into the flowfield. The second ignition occurs just downstream of the wedge turn angle and is due to shock-induced combustion caused by the compression of the two OSWs along the wedge surface. The structure of the induction region in Fig.~\ref{fig:double-angle_inviscid} and the OSW-to-ODW transition at 30 $\mu$s resembles the smooth transition that occurs for higher Mach number flows. Previous studies \cite{teng2017initiation} have shown this to be characterized by the weakening of the reflected secondary shock (present in abrupt transitions) to the point of vanishing. This is seen in comparing the steady-state temperature at 30 $\mu$s in Fig.~\ref{fig:double-angle_inviscid} to the that of the abrupt transition in Fig.~\ref{fig:m7_close}.

\subsection{Viscous wedge surface}

	The double-angle wedge is used in cases with a viscous, no-slip wedge surface condition. The inflow is held at Mach 5, attainable in the CATER experimental facility at UCF. The inflow temperatures considered are 600 K, 700 K, and 800 K. The previous 800 K cases in Figs.~\ref{fig:ucf_m7} and \ref{fig:double-angle_inviscid} suggest that these low Mach number flows will not ignite. Nonetheless, the presence of a boundary layer along the viscous surface is found to have significant impact on the nature of the flowfield.

\subsubsection{Stable States}

\begin{figure}
		\centering
		\includegraphics[width=88mm]{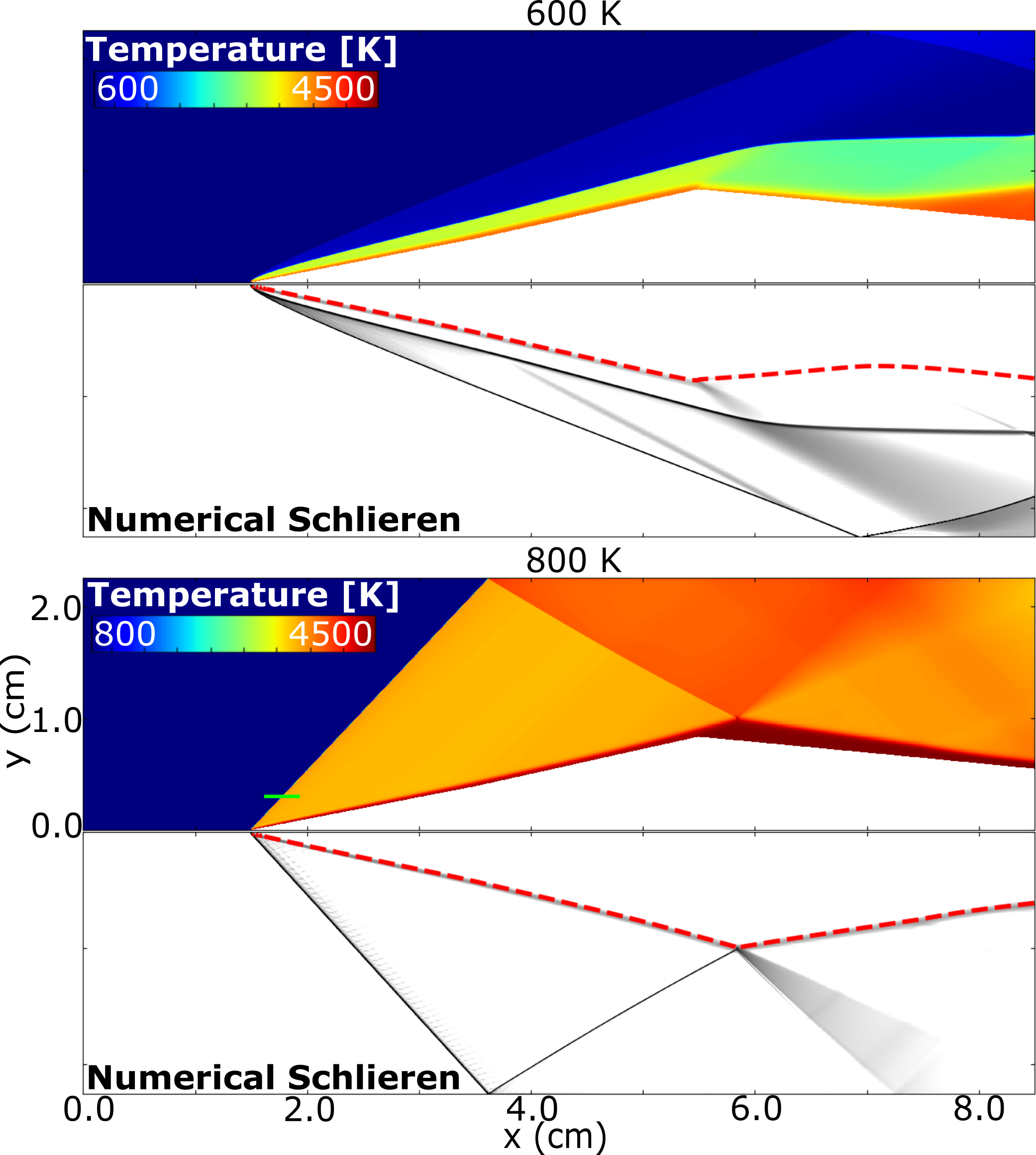}
		\caption{Steady-state flowfields for the double-angle wedge in 600 K (top) and 800 K (bottom) freestreams. The green line denotes the slice along which the quantities in Fig.~\ref{fig:props} are taken.}
		\label{fig:600_800}
	\end{figure}

	The inflow temperatures of 600 K and 800 K produce the steady-state flows shown in Fig.~\ref{fig:600_800}, with the schlieren plots mirrored vertically and shown underneath the temperature contours. Again, the schlieren is overlaid by a single contour for the sonic line.

	The 600 K flow ignites in the boundary layer along the wedge surface, where the no-slip condition is enforced, due to the elevated temperatures of the stagnated flow along the adiabatic surface. The burning boundary layer augments the angle of the OSW. Even with the increased shock strength, the unburned post-shock flow does not autoignite within the flow residence time. Without autoignition, the flow reaches a steady state without a detonation forming.

\begin{figure}
		\centering
		\includegraphics[width=88mm]{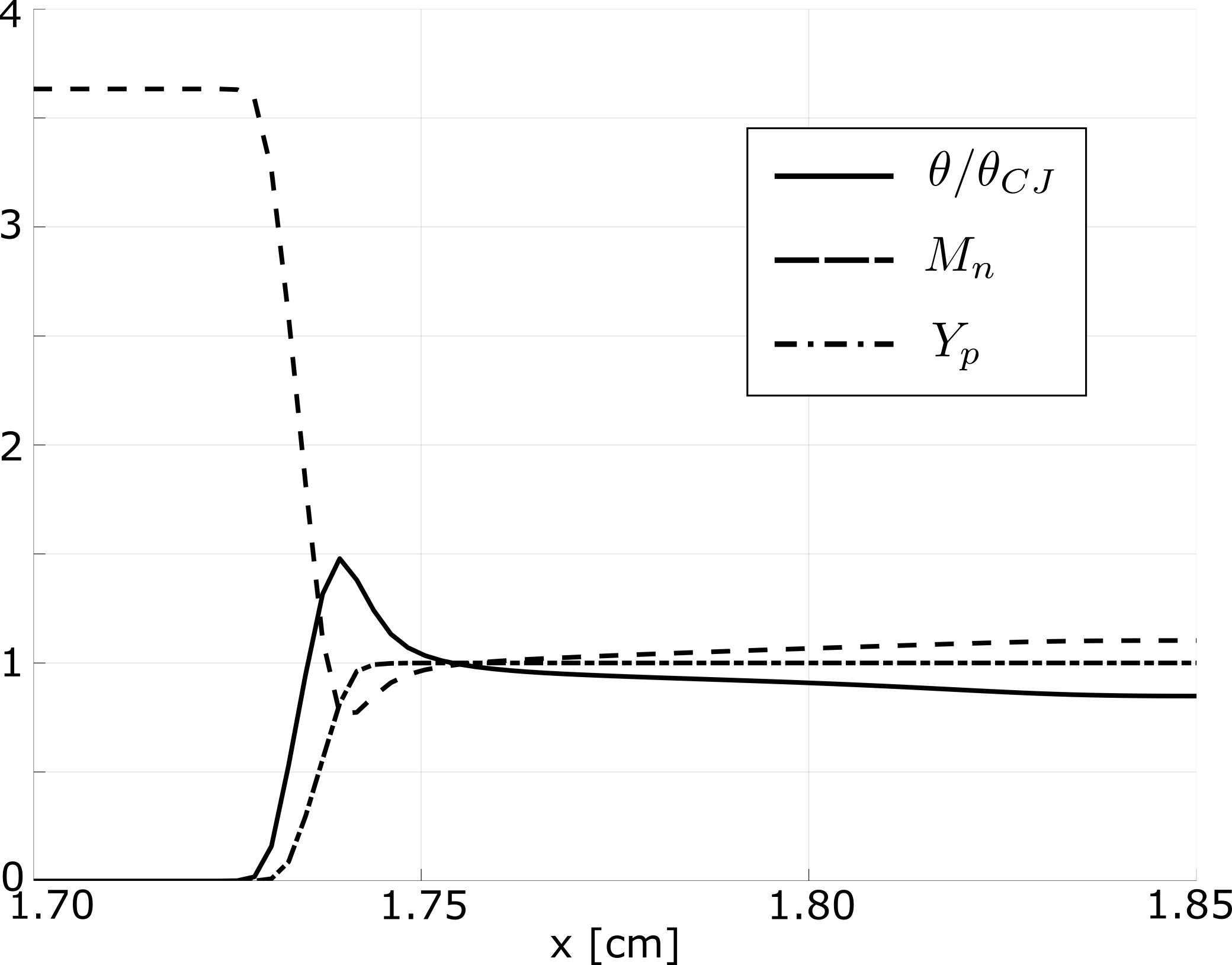}
		\caption{Flow angle normalized by $\theta_{CJ}$, normal component of Mach number, and product mass fraction across the stabilized ODW for the 800 K inflow over the viscous double-angle wedge.}
		\label{fig:props}
	\end{figure}

	The 800 K case, however, forms a prompt ODW at the leading edge. Immediate autoignition occurs because of the increased pressure and temperature jump near the leading edge as the boundary layer develops. The ODW penetrates into the freestream, growing at a constant wave angle until it reflects from the top wall, as shown. This development seems to be unaffected by the small flow disturbance resulting from the second angle along the double-angle wedge. The stabilized ODW angle is about 46.6$^o$, which is precisely equal to $\beta_{CJ}$ predicted by the stability analysis for a Mach 5, 800 K flow. The flow deflection angle is equal to $\theta_{CJ}$ just behind the ODW, at a distance that coincides with the approximate point of complete burning, but the flow angle relaxes downstream to the wedge surface angle. Likewise, the normal component of the Mach number reaches unity at precisely the same point of complete burning, but increases slightly downstream. Flow properties along the green line in Fig.~\ref{fig:600_800} are plotted in Fig.~\ref{fig:props}. The flow angle, $\theta$, normalized by the predicted $\theta_{CJ}$, the component of the Mach number normal to the ODW, $M_n$, and the product mass fraction, $Y_p$, all reach unity at approximately the same point downstream of the ODW. As previously stated, the flow angle relaxes to near that of the wedge surface, resulting in the continued increase of $M_n$ past the point of unity.

\subsubsection{Quasi-stable State}

\begin{figure*}[h]
		\centering
		\includegraphics[width=7in]{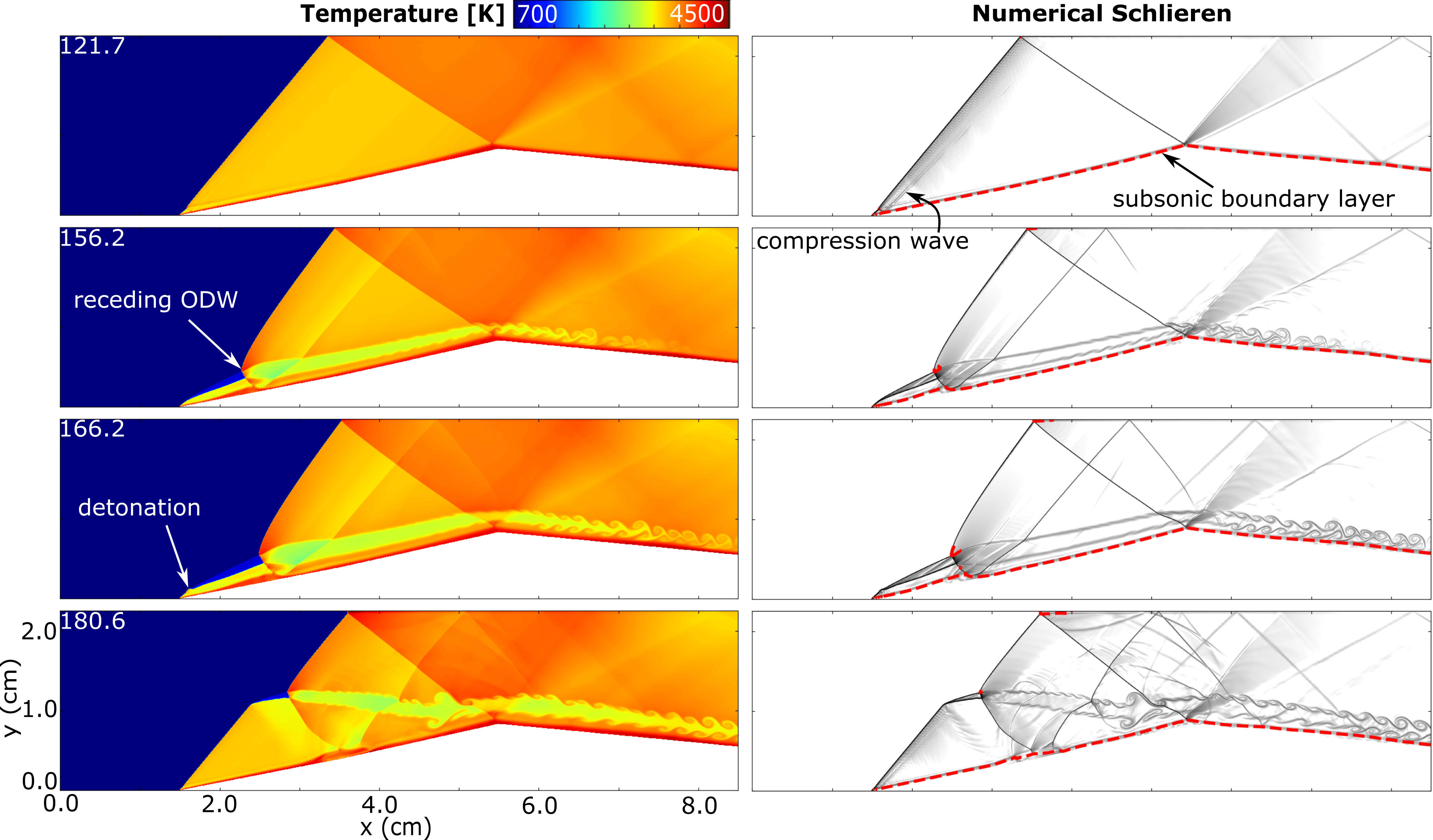}
		\caption{Viscous double-angle wedge in a Mach 5, 700 K flow. The dashed line encloses the subsonic region, and timestamps are given in $\mu$s.}
		\label{fig:700K_odw_evol}
	\end{figure*}

	An inflow temperature of 700 K for the viscous double-angle wedge yields a unique quasi-stable mode. Figure~\ref{fig:700K_odw_evol} shows the evolution of this case, starting from a point which resembles the stable structure of the previously described 800 K case in Fig.~\ref{fig:600_800}. A prompt ODW forms, ignited in the boundary layer at the leading edge, and grows at a constant angle into the freestream. The angle of growth is equal to the predicted $\beta_{CJ}$ for the inflow conditions, and $\theta_{CJ}$ is reached behind the ODW at a distance coinciding with complete burning and the point at which $M_n$ is unity. This case differs from the high temperature 800 K case in Fig.~\ref{fig:600_800} case in that it contains a small induction region, evidenced by a compression wave emanating from near the leading edge and running slightly divergent to the ODW, labeled in the schlieren contour at 121.7 $\mu$s. This compression wave is the reflection of the induction region's terminating reaction front from the wedge surface, and is present in the inviscid wedge surface cases as well. 

	While the established detonation is a CJ-ODW, the reaction front in the induction region is underdriven, which causes the entire ODW structure to recede. The induction region grows, producing a structure that resembles the typical inviscid wedge surface cases, with the addition of a burning boundary layer, as seen at time 156.2 $\mu$s in Fig.~\ref{fig:700K_odw_evol}. This boundary layer augments the OSW, increasing the wave angle above what would result from the equivalent inviscid wedge surface. Before the receding ODW can stabilize, a secondary detonation is initiated near the leading edge, at 166.2 $\mu$s. This triggers a CJ-ODW to penetrate the flow again and establish the same structure that was initially produced. This process of receding ODW followed by reignition near the leading edge repeats with a period of about 100 $\mu$s, or a frequency of 10 kHz. Proposed mechanisms of this periodicity are discussed in Secs.~\ref{bl} and \ref{red mech}.

\section{Discussion}
\label{Discussion}

\subsection{Inviscid ignition criterion}
\label{ign crit}

	Previous studies \cite{li1994detonation,morris1998shock,silva2000stabilization} have confirmed that initiation of an ODW on an inviscid wedge surface depends on shock-induced combustion by the leading OSW that forms on the wedge. Ignition, therefore, occurs downstream of the leading edge and the flow residence time on the wedge surface must exceed the ignition delay time of the flow behind the OSW. The inviscid wedge surface cases considered here exhibited the same behavior and, most notably, certain cases which lie inside the traditional stability limits shown in Fig.~\ref{fig:inv_stab} did not ignite along the surface. Such was the case for the 500 K, Mach 7 inflow with a 20$^o$ wedge, shown in Fig.~\ref{fig:m7_odw}. Given that the traditional stability theory does not take into account chemical kinetic timescales, an additional ignition criterion can be used to modify the stability plots to account for ignition delay and wedge surface length.

\begin{figure*}[h]
		\centering
		\includegraphics[width=6.5in]{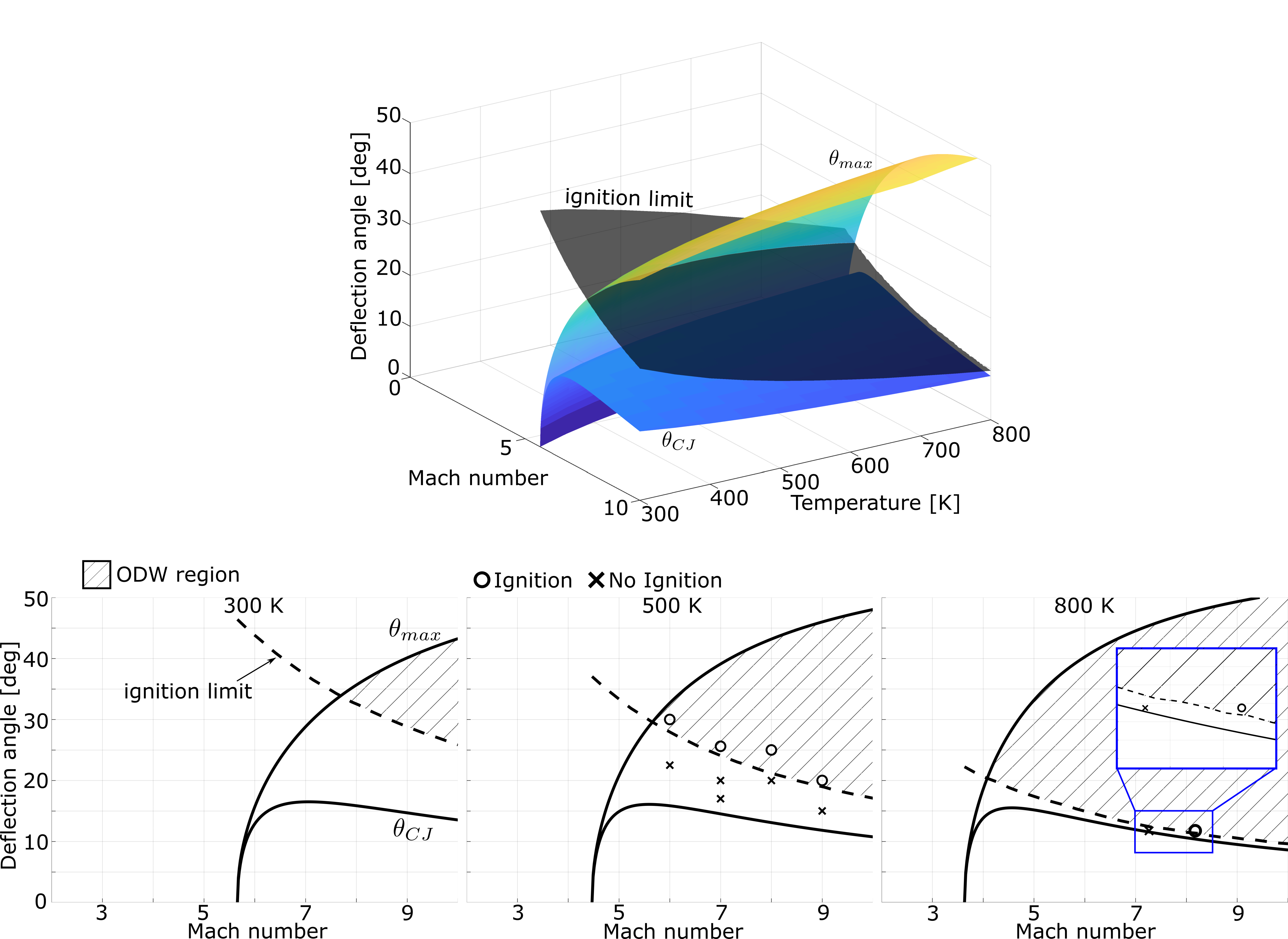}
		\caption{Ignition and stability limits for an ODW, modified by ignition delay behind OSW on a wedge of length 4 cm.}
		\label{fig:surf}
	\end{figure*}

	Based on predicted ignition delay times over a range of post-shock conditions, it was determined whether or not ignition would occur along a wedge of a known length. These results were used to develop a new curve to be overlaid on the traditional stability map. Figure~\ref{fig:surf} shows the modified stability map over a range of Mach numbers and temperatures for a wedge surface of length 4 cm. For inflow Mach number and wedge deflection angles that fall below the new curve, ignition will not occur on the wedge surface and an ODW cannot form. Fig.~\ref{fig:surf} shows several two-dimensional slices of the three-dimensional map at inflow temperatures of 300 K, 500 K, and 800 K. The new ignition criterion is plotted as a dashed line, and the reduced regions of stable ODW production are shaded.

	Additional data points are included on the 500 K plot for each of the simulations performed using domain 1. The cases in which ignition occurred, and an ODW formed, are plotted as circles, and those that did not ignite are plotted as x's. The ignition criterion is seen to successfully predict the formation of an ODW. The cases presented in Figs.~\ref{fig:m7_odw}-~\ref{fig:m6_t30_evol} make up three of these data points, and their locations relative to the ignition criterion explains the presence or lack of ignition.

	Likewise, the slice at 800 K includes data points showing the results of the two cases performed using domain 2. The double-angle wedge is represented as a single data point with deflection angle equal to the average of the two angles, 11.75$^o$. The Mach 7.3 inflow lies within the traditional stability limits, but below the additional ignition criterion, explaining the lack of ignition observed in Fig.~\ref{fig:ucf_m7}. An inflow of Mach 8.2 barely meets the ignition criterion, which explains the location of the terminating reaction front at the end of the wedge surface in Fig.~\ref{fig:double-angle_inviscid}.

	With the new ignition criterion, the formation of an ODW on an inviscid wedge surface is highly dependent on the wedge surface length. In all inviscid wedge surface cases that ignite, the length of the induction region precisely corresponds to a flow residence time that is equal to the ignition delay for the post-shock conditions. Making the wedge length shorter or longer moves the ignition criterion curve up or down, respectively.

\subsection{Effect of boundary layers}
\label{bl}

\begin{figure}[h]
		\centering
		\includegraphics[width=88mm]{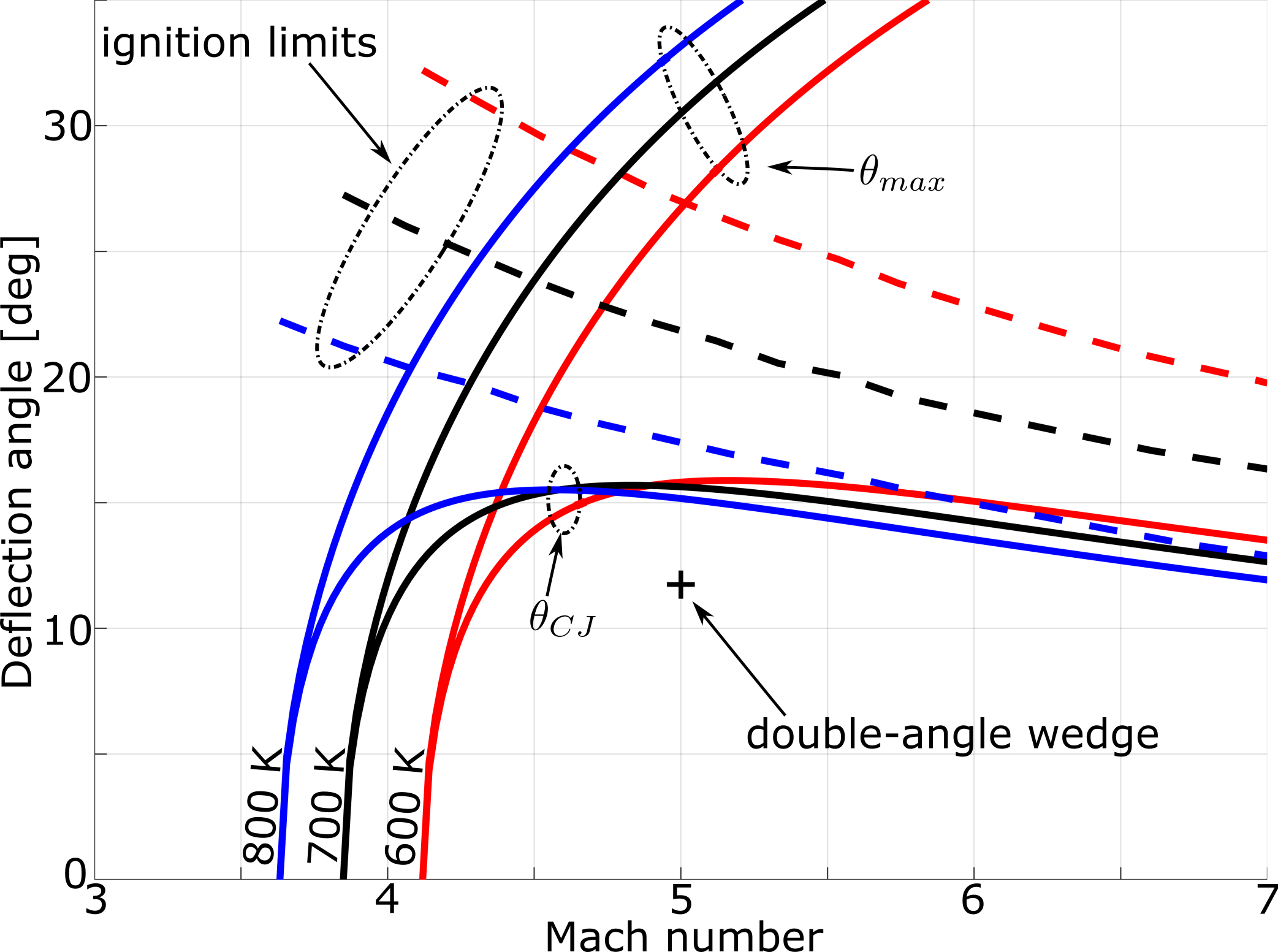}
		\caption{Traditional stability predictions (solid) and ignition limits (dashed) for the double-angle wedge (+) in a Mach 5 flow at 600 K (red), 700 K (black), and 800 K (blue).}
		\label{fig:678stab}
	\end{figure}

	The stability curves and ignition criterion curves are shown in Fig.~\ref{fig:678stab} for the three inflow temperatures simulated in domain 2 with a no-slip condition on the wedge surface. The average of the two angles on the double-angle wedge is also shown on the figure. For all three static temperatures in a Mach 5 flow, the double-angle wedge lies below the lower stability limit, $\theta_{CJ}$, and well below the ignition criterion. Nonetheless, when a boundary layer is present, the 700 K and 800 K flows detonate, as shown in Figs.~\ref{fig:600_800} and ~\ref{fig:700K_odw_evol}. Moreover, the 800 K flow over an inviscid double-angle wedge required an inflow Mach number greater than 8 to detonate (shown in Fig.~\ref{fig:double-angle_inviscid}), but the viscous wedge surface enabled ODW formation for a Mach number as low as 5. While the inviscid wedge surface cases demonstrate the predictability of ODW initiation and stability based on the ignition criterion, these viscous wedge surface cases suggest significant augmentation by the presence of a boundary layer. 

\begin{figure}[h]
		\centering
		\includegraphics[width=88mm]{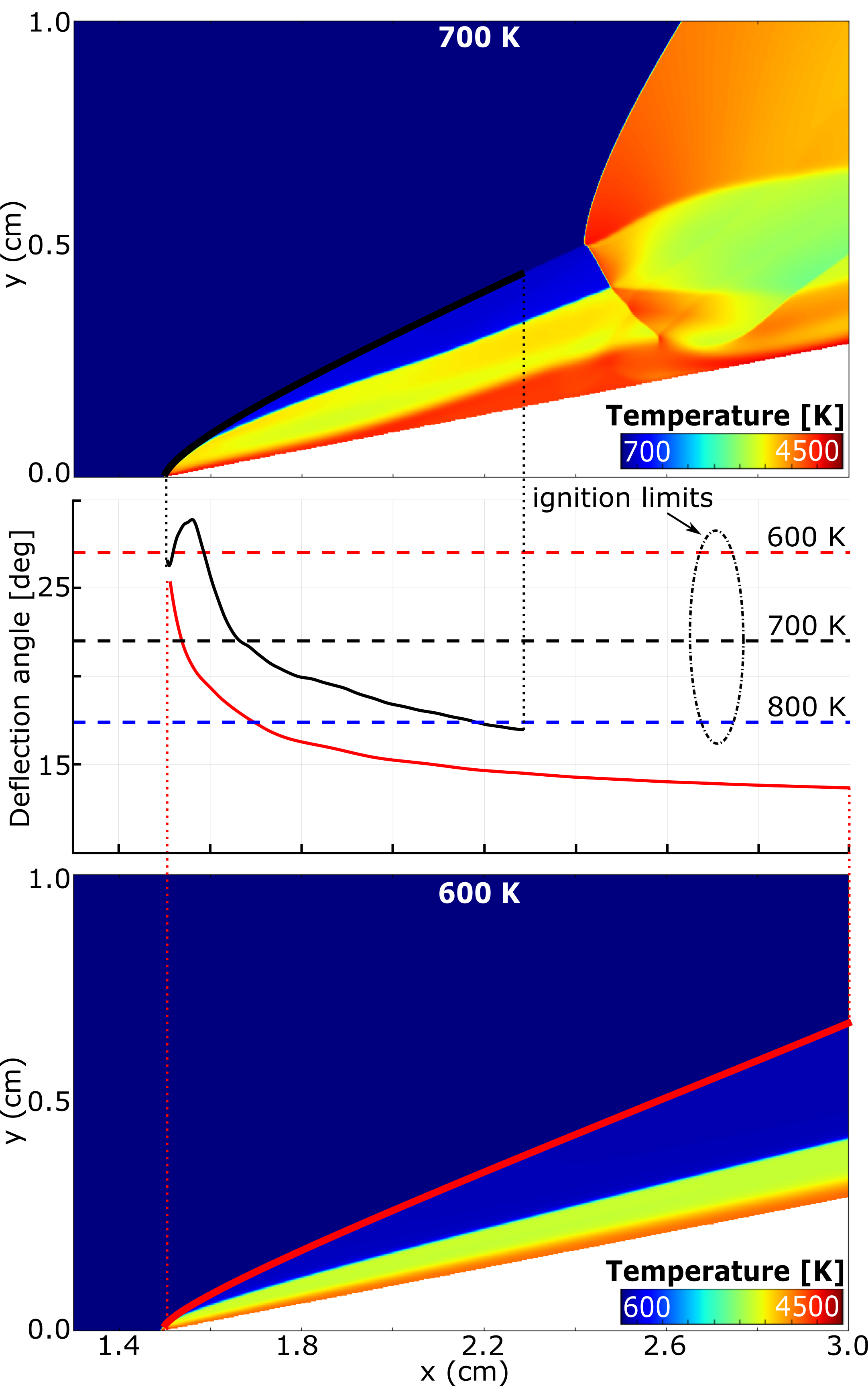}
		\caption{Local deflection angles along the augmented OSWs for the transient 700 K (black) and steady-state 600 K (red) flows.}
		\label{fig:deflection}
	\end{figure}

	The role of the boundary layer is examined through an analysis of the augmented deflection angle imposed by the boundary layer, in light of these ignition criteria. Figure~\ref{fig:deflection} shows the 700 K flow (top) during the transient recession of the ODW, and the 600 K flow (bottom) at steady state. These are the viscous wedge surface cases in domain 2 shown in Figs.~\ref{fig:600_800} and \ref{fig:700K_odw_evol}. The post-shock deflection angle is shown by the solid lines (middle) for each of these cases, plotted along the OSW from the leading edge to some distance downstream. The value of the deflection angle at the ignition criterion for the Mach 5 flow is shown by the dashed lines. While the leading edge deflection angle on the inviscid wedge surface would be equal to the 11$^o$ wedge angle, the deflection angle on the viscous wedge surface varies between the 600 K and 700 K temperatures and along inflowing streamlines. The burning boundary layer blunts the leading edge, increasing the deflection angle significantly in that region, but the shock wave angle and deflection angle are relaxed downstream. The comparison of the local augmented deflection angles to the ignition limits explains the behavior seen in each of these cases.

	The deflection angle produced by the 600 K flow peaks near the leading edge at about 25$^o$. Thus, the deflection angle still does not exceed that which would provide sufficient shock-compression to ignite the supersonic flow along the wedge surface. Therefore, the flow reaches the steady state shown, the burning boundary having been ignited by viscous heating and local flow stagnation. 

	For the 700 K inflow, the maximum augmented deflection angle near the leading edge exceeds the ignition limit of 22$^o$. The post-shock ignition delay for the maximum deflection angle of about 29$^o$ corresponds to an induction length of about  2.8 mm. Moving downstream along the OSW, the deflection angle decreases and drops below the ignition limit within 2 mm of the leading edge. The inflowing streamlines deflected by an amount sufficient to ignite the mixture are consumed by the deflagration front along the burning boundary layer, and therefore are not given sufficient time to autoignite. Meanwhile, the streamlines deflected by an angle below the ignition limit remain unburned by the deflagration, but do not have a sufficiently short ignition delay time to ignite. This explains why the receding ODW within the induction region is underdriven, as noted in the description of Fig.~\ref{fig:700K_odw_evol}. 

	From the start of this 700 K case, the early development of the burning boundary layer allows the autoignition of the maximally deflected streamlines, enabling the initial ODW growth. A small induction region forms and, as the boundary layer develops and grows, these maximally deflected streamlines are consumed by the deflagration within the induction region. This causes the ODW within the induction region to be underdriven and recede. The transient nature of this particular case gives rise to redetonations, the mechanism of which is discussed in Sec.~\ref{red mech}.

	The case with an 800 K inflow does not render itself to the same analysis since it immediately detonates and does not exhibit a transient induction region to consider. Still, the analysis of the 600 K and 700 K inflows suggests that the augmentation of the deflection angle by the boundary layer is sufficient to cause autoignition during boundary layer development, and for this prompt ODW structure to remain stable. The ignition limit for an 800 K inflow is included in Fig.~\ref{fig:deflection} to show the relatively small increase in deflection angle that is required. 

\subsection{Redetonation mechanism}
\label{red mech}

\begin{figure*}
		\centering
		\includegraphics[width=7in]{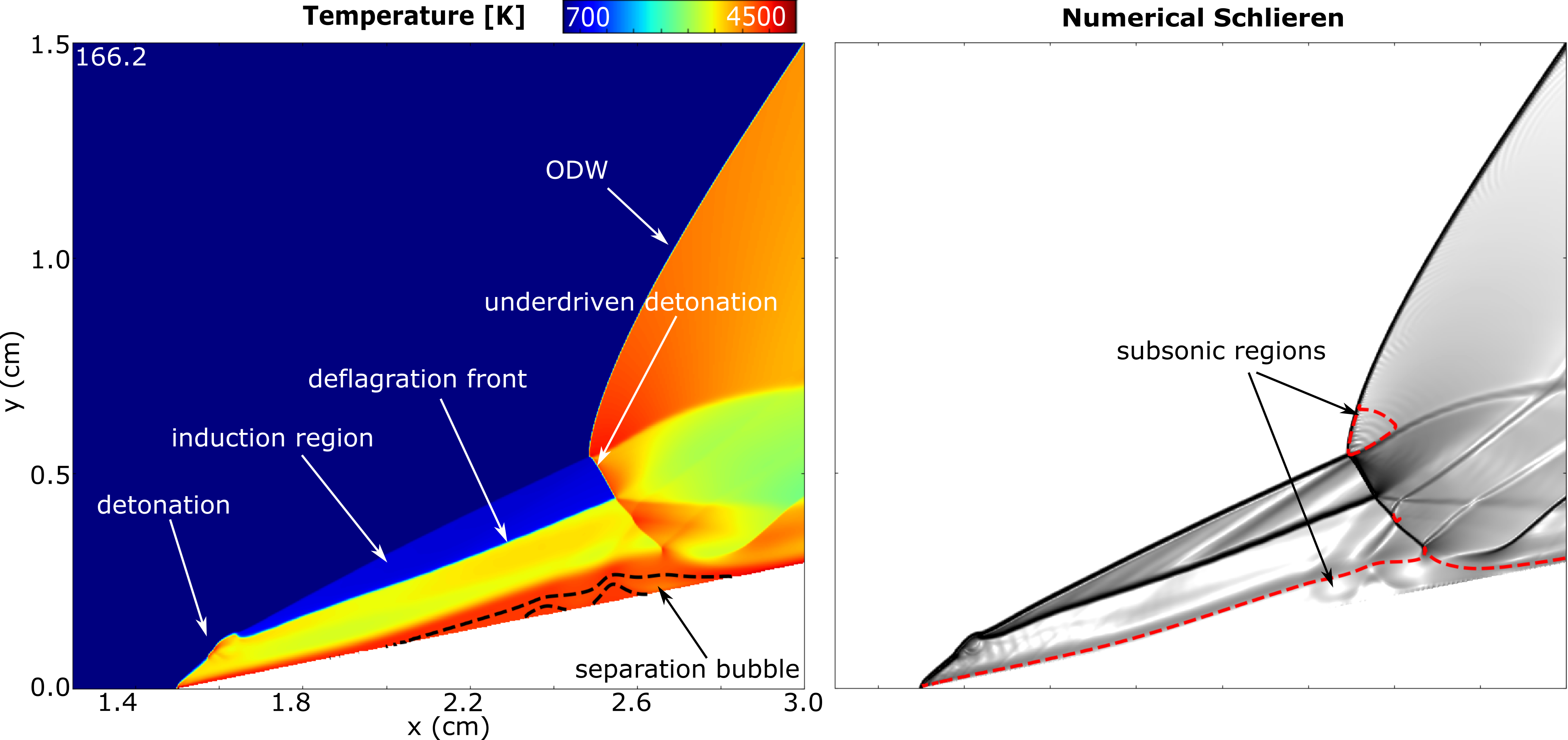}
		\caption{Transient ODW structure and redetonation for Mach 5, 700 K flow at 166.2 $\mu$s.}
		\label{fig:700K_zoom}
	\end{figure*}

	The mechanism driving the periodicity in the quasi-stable, 700 K viscous wedge surface case is illuminated by Fig.~\ref{fig:700K_zoom}, where a close-up of the leading edge is shown at the point of redetonation. This coincides with 166.2 $\mu$s in Fig.~\ref{fig:700K_odw_evol}. The growing induction region is terminated by an underdriven detonation wave that is approximately normal to the local flow. A portion of the ODW remains overdriven, denoted by the subsonic region in the schlieren. The other subsonic region along the wedge surface grows as the structure recedes and the terminating detonation intersects the boundary layer in a complex shock structure. This gives rise to a separation bubble, the boundaries of which are designated by the black dashed line marking the contour where the x-component of velocity is equal to zero. The shock wave impinging on the reactive boundary layer, and subsequent growth of the separation bubble, imposes a growing back-pressure on the subsonic portion of the boundary layer. As the downstream perturbations and back-pressure grow, the leading edge shock is forced to strengthen, eventually perturbing the deflagration front to the point of detonation. This cycle continues indefinitely.

\subsection{Resolution study}
\label{resolution}

\begin{figure}[h]
		\centering
		\includegraphics[width=88mm]{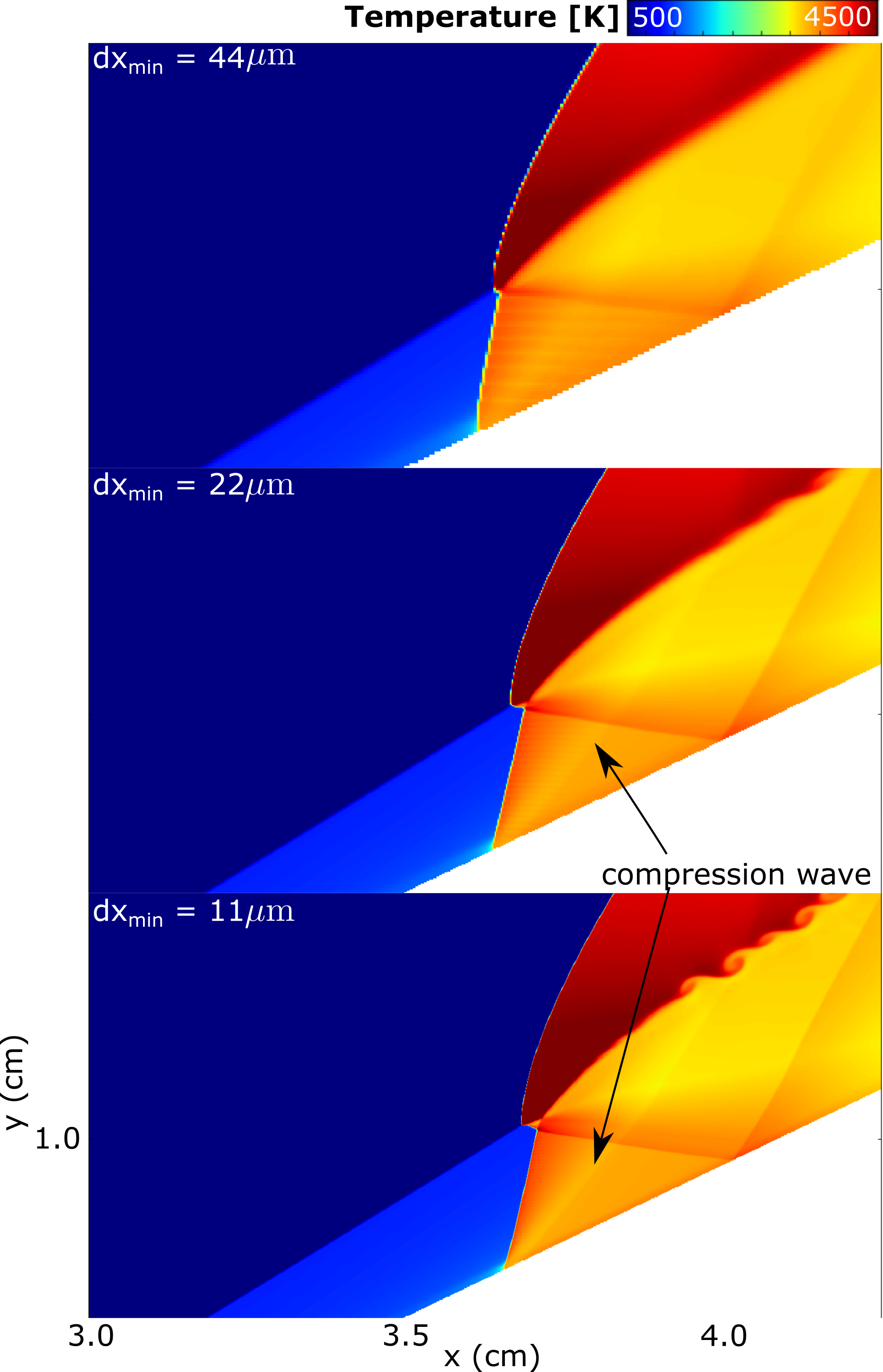}
		\caption{Resolution study performed using a 25.6$^o$ wedge in a Mach 7, 500 K flow.}
		\label{fig:resolution}
	\end{figure}

	A resolution study was performed to determine the level of computational grid refinement necessary to capture the flow features of interest. The case investigated is the 500 K, Mach 7 inflow in domain 1 with a wedge angle of 25.6$^o$ and a wedge length of 4 cm. This is the same as presented in Figs.~\ref{fig:m7_odw} and \ref{fig:m7_close}. Three grid resolutions were used, each with $dx_{max}$ = 1.4 mm, and $dx_{min}$ = 44, 22, and 11 $\mu$m. The simulation was carried to a steady state for each resolution. The resulting ODW structures are shown in Fig.~\ref{fig:resolution}. 

	While the termination of the induction region in each case occurs at approximately the same distance from the leading edge, the shape of the terminating wave front (determined to be a CJ-ODW) appears slightly bowed in the least refined case. This suggests that the strengthening of this reaction front into a CJ-ODW is less defined. Further evidence of this can be seen in comparing the compression waves emanating from the ignition point on the surface at a shallower angle than the CJ-ODW. This compression wave, labeled in the more refined grids, turns the flow near the surface back to tangency, and is also responsible for the terminating reaction front becoming pressure driven. At lower refinement this compression is smoothed to the point of being hardly noticeable, causing the structure of the CJ-ODW to become bowed.

	The second-most refined case, $dx_{min}$ = 22 $\mu$m, contains this compression wave and the same steepening of the CJ-ODW as in the most refined case, $dx_{min}$ = 11 $\mu$m. These two resolutions also have $\lambda$ structures with comparable definition. Specifically, the upper branch of the $\lambda$ shape is more clear in the middle and high resolution cases than in the case with lowest resolution. While the flow instabilities along the slip-line downstream of the ODW are not as developed and defined in the middle resolution as in the highest, this instability is not a driving factor behind the ODW initiation and stability, which are the physics of interest in the present study. Therefore, the second-highest resolution, with $dx_{min}$ = 22 $\mu$m, was determined to be sufficient for the present calculations.

\section{Conclusions}
\label{Conclusions}

	The ignition and stability of wedge-stabilized oblique detonation waves (ODWs) and the effect of a boundary layer along the wedge surface were studied using multidimensional, unsteady numerical simulations of supersonic flows of stoichiometric hydrogen-air mixtures. This paper compared a range of inviscid wedge surface cases, varying inflow Mach number, static temperature, and wedge geometry. Several viscous wedge surface cases were discussed, using an inflow at Mach 5 and temperatures of 600 K, 700 K, and 800 K.

	The results of the inviscid wedge surface cases are consistent with existing literature \cite{li1993effects, silva2000stabilization, liu2015analytical, teng2014numerical} in terms of the smooth or abrupt induction region structures seen at relatively high and low Mach numbers, respectively, and in particular transient cases leading to prompt ODWs forming at the leading edge. The transience or stability is explained by the overdriven or underdriven nature of the ODWs that are formed within the induction region. A new ignition criterion curve was added to the traditional stability limits in order to include the effects of chemical kinetic timescales and wedge surface length on the prediction of ignition and formation of ODWs. This ignition criterion was demonstrated to accurately predict formation and ignition location of an ODW on an inviscid wedge surface.

	The viscous wedge surface simulations were performed using a double-angle wedge proposed to be used in experiments at the Center for Advanced Turbomachinery and Energy Research (CATER) at University of Central Florida (UCF). While the application of the established ignition criteria to the wedge geometry predicts no ODW formation, the presence of the boundary layer augments the oblique shock wave (OSW) angle and provides sufficient shock compression for the 700 K and 800 K Mach 5 flows to ignite and form ODWs. Thus, boundary layers are shown to have a significant effect on ODW initiation and structure. The 800 K flow ignites and reaches a steady-state, prompt ODW at the leading edge, whereas the 700 K flow ignites and exhibits a unique oscillatory state of ODW recession followed by redetonation near the leading edge. Mechanisms of this periodic process were proposed, the driving factors being the underdriven ODW terminating the small induction region, and a subsequent separation bubble imposing an increasing backpressure on the subsonic portion of the boundary layer. The 600 K flow does not detonate but reaches a steady state with a burning boundary layer since the OSW, even though significantly augmented by the boundary layer, does not provide sufficient flow deflection to ignite the mixture.

	Future work will investigate and characterize supersonic burning boundary layers in terms of growth rates and thickness to better predict the augmentation of a leading edge OSW. Methods of prediction will be applied to the ignition criterion presented in the present work to make ODW formation and stability predictions more robust and accurate, and enable these predictions to be performed without the need to run expensive numerical simulations. 

\section*{Acknowledgments}
\label{Acknowledgments}

	This work was supported by the Base Program at the Naval Research Laboratory, provided through the Office of Naval Research. The authors thank Professor Ryan Houim at the University of Florida for the use of HyBurn, the computational fluid dynamics code used to perform the simulations described in this work. The computations were performed on a Department of Defense High Performance Computing Center System at the Army Research Laboratory. The authors also thank Professor Kareem Ahmed at the University of Central Florida for providing the double-angle wedge geometry for consideration in this work.

\bibliography{Manuscript} 
\bibliographystyle{elsarticle-num}


\end{document}